\documentclass[aps,twocolumn,pre]{revtex4}
\usepackage{epsfig}
\newcommand{\n}{\noindent}

\begin{document}

\title{Nodal domain distribution of rectangular drums \footnote{\it Presented
in National Conference on Nonlinear Systems and Dynamics (Feb. 24-26, 2005),
Aligarh Muslim University, India.}}
\author{U. Smilansky and R. Sankaranarayanan\footnote{Present Address:
Centre for Nonlinear Dynamics, Department of Physics, Bharathidasan
University, Tiruchirappalli 620024, India.}}
\affiliation{Department of Physics of Complex Systems \\ Weizmann Institute
of Science, 76100 Rehovot, Israel.}

\begin {abstract}
We consider the sequence of nodal counts for eigenfunctions of the
Laplace-Beltrami operator in two dimensional domains. It was conjectured
recently that this sequence stores some information pertaining to the
geometry of the domain, and we show explicitly that this is the case for
the family of rectangular domains with Dirichlet boundary conditions.
\end {abstract}

\maketitle

\section{introduction}
One of the major problems in mathematical physics is concerned
with the geometrical information stored in the spectrum of the Laplace
Beltrami operator
\begin{equation}
-\triangle\psi_j({\bf r}) = E_j\psi_j({\bf r}); \;\; {\bf r}
\in \Omega(\alpha) \ .
\end{equation}

\n The spectrum is ordered such that $E_{j-1}\ \le E_j\le E_{j+1}$ and
$\Omega(\alpha)$ is a connected compact region, parameterized by $\alpha$,
on a 2D Riemannian manifold. If $\Omega(\alpha)$ has a boundary, Dirichlet
boundary conditions are assumed. The corresponding physical system could be
a vibrating drum. In 1911 H. Weyl showed that the number of eigenvalues up
to energy E is
\begin{equation}
N(E) \sim {AE\over 4\pi}, \;\;\;\; \hbox{as} \;\;
E\rightarrow \infty
\end{equation}

\n where $A$ is the area of $\Omega$. Subsequent research have shown
(see e.g., \cite{clark67}) that each of the terms in the asymptotic series
of $N(E)$ provides further geometrical information on the boundary. This
prompted M. Kac to ask, `can one hear the shape of a drum ?' \cite{kac66}.
That is, `is it possible to uniquely define the shape of the drum from
the spectrum ?' It is known by now that for certain classes of domains
the answer to Kac's question is positive, whereas there exists a large
set of {\it isospectral} domains which are not {\it isometric}.
(Ref. \cite{zeldich} gives an updated review of this subject.)

In the present note we would like to investigate the geometrical information
stored in yet another sequence of numbers which are derived from the
eigenfunctions $\psi_j$. Considering real eigenfunctions $\psi_j$, we count
the number $\nu_j$ of {\it nodal domains} which are the connected domain
where $\psi_j$ has a constant sign. The nodal domains are separated by
the {\it nodal lines} where $\psi_j=0$. The sequence 
$\left \{\nu_j\right\}_{j=1}^{\infty}$ is the sequence of nodal counts.
According to Courant's Nodal theorem $\nu_j \le j$. This fundamental
theorem reveals the deep connection between the spectrum and the nodal
count. It is convenient to define the {\it normalized} nodal domain
numbers $\xi_j = \nu_j/j$. Because of Courant's theorem 
$0 \le \xi_j \le 1$. This estimate has been further refined (for domains
in $\mathbf{R}^2$ ) \cite{pleijel56}
\begin{equation}
\limsup_{j\rightarrow\infty}\; \xi_j = 0.691 \ldots
\end{equation}

\n Following \cite{uzy02}, we study the distribution of the normalized nodal
numbers in the spectral interval $I=[E^0,E^1]$
\begin{equation}
P(\xi,I) = {1\over N_I}\sum_{E_j\in I} \delta(\xi-\xi_j)
\label{dbn}
\end{equation}

\n where $N_I$ is the number of levels in the interval $I$.

In Ref. \cite{uzy02} the above distribution has been introduced as a tool
to distinguish between systems which are integrable (separable) or classically
chaotic. For the class of separable domains, it was shown that the limit
distribution
\begin{equation}
P(\xi) = \lim_{E\rightarrow\infty} P(\xi,I)
\end{equation}

\n exists. This has universal features: ({\it a}) there exists a system
dependent parameter $\xi'$, maximum value of the nodal domain number, such
that $P(\xi)=0$ for $\xi > \xi'$ and ({\it b}) for $\xi\approx\xi'$, 
\begin{equation}
P(\xi) = {C\over \sqrt{1 -\xi/\xi'}} \, .
\label{uexp}
\end{equation}

\n The constant $C$ is system dependent, but the order of the singularity
is universal and depends only on the dimensionality. (It was recently shown
that the exponent for domains in $d$ dimensions  is $(d-3)/2$.)

The dependence on the geometry of the domain can come only through the
parameters $\xi',C $  or the details of the function $P(\xi)$ away from
the universal domain. Indeed, the limiting distributions for the rectangular
and circular boundaries were computed in \cite{uzy02} and found to be
different as expected. However, as will be shown below, the function
$P(\xi)$ does not distinguish between different rectangles. That is
$\xi' = 2/\pi$ and 
\begin{equation}
P(\xi) = {\left[1 - {(\pi\xi/2)}^2 \right]}^{-1/2}
\end{equation}

\n for all rectangles! Note that for $\xi\approx 2/\pi$, this result
coincides with the universal expression (\ref{uexp}) with $C=1/\sqrt 2$.
The new result of the present note is that the dependence of $P(\xi,I)$
on the {\it finite} spectral interval $I$ contains sufficient information
to resolve between different rectangles. Thus, by counting nodal domains
one can deduce the shape of the (rectangular) drum. It should be emphasized
at the outset that the nodal count sequence involves dimensionless integers,
and therefore it cannot provide any scale information. Hence, when we say
``resolve'' we mean ``resolve up to a scale''.

\section{rectangles}

We consider the Dirichlet spectrum of a domain bounded in a rectangle with
sides $L_x$ and $L_y$. Denoting $\alpha = L_x/L_y$ and choosing $L_x=\pi$,
the spectrum is given by
\begin{equation}
E = n^2 + \alpha^2m^2\ ,
\end{equation}

\n where $n,m=1,2,3\ldots$ and $0 < \alpha < 1$. Since the system is separable
in rectangular co-ordinates the nodal domain number is simply $\nu_j=n m$,
and $j=N(n^2 + \alpha^2m^2)$ where $N(E)$ is the spectral counting function.
The leading terms in the asymptotic expansion of $N(E)$ are
\begin{equation}
N(E) \simeq {1\over 4\pi} \Big[AE - L\sqrt{E}\Big] \label{weyl}
\end{equation}

\n where $A,L$ are the area and perimeter of the boundary
respectively \cite{morse}. In terms of $\alpha$,
\begin{equation}
N(E) \simeq {\pi E\over 4\alpha}
\left(1-{2\over\pi}{1+\alpha\over\sqrt{E}}\right)\, .
\end{equation}

\n Introducing the transformation
\begin{equation}
n(E,\theta) = \sqrt{E}\cos\theta \;\; ; \;\;
m(E,\theta) = \sqrt{E}\sin\theta/\alpha \, ,
\end{equation}

\n the normalized nodal-domain number can be approximated by
\begin{equation}
\xi_j(E,\theta) = {2\over\pi} \sin 2\theta \left[1 - {2\over\pi}
{(1+\alpha)\over\sqrt{E}}\right]^{-1} \, .
\end{equation}

\n Converting the summation in eq.(\ref{dbn}) into an integral, we obtain
the leading terms in the asymptotic expansion of $P(\xi,I)$ in the large
$E$ limit
\begin{equation}
P(\xi,I) \simeq {1\over 2\alpha N_I}
\int_{E^0}^{E^1}\int_0^{\pi/2} \delta
\Big[\xi - \xi_j(E,\theta)\Big]\;dE\;d\theta
\end{equation}

\n where
\begin{equation}
N_I \simeq {\pi\over 4\alpha}\left\{(E^1-E^0)-{2\over\pi}
(1+\alpha)\left(\sqrt{E^1}-\sqrt{E^0}\right)\right\} \, .
\end{equation}

\n Introducing the variable $x=\sqrt{E/E^0}$
\begin{equation}
P(\xi,I) = {E^0\over\alpha N_I} \int_1^g\int_0^{\pi/2} x\;\delta
\left[\xi - {2\over\pi}{\sin 2\theta\over(1-\epsilon/x)}\right] \;
dx\; d\theta
\end{equation}

\n where
\begin{equation}
g=\sqrt{E^1\over E^0}, \;\;  \epsilon(\alpha) =
{2\over\pi}{(1+\alpha)\over\sqrt{E^0}}.
\end{equation}

\n The integral reduces to
\begin{equation}
P(\xi,I) = {E^0\over\alpha N_I} \int_1^l x {\left[{2\over\pi}
{\cos 2\theta_0\over(1-\epsilon/x)}\right]}^{-1} \; dx
\end{equation}

\n where $\sin 2\theta_0 = {\pi\xi\over 2} \left(1-{\epsilon\over x}\right)$
and
\begin{equation}
l = \left\{
\begin{array}{ll}
g,&\hbox{if} \;\; \xi < {2\over\pi} \\
\hbox{min}\left[g,\epsilon{\pi\xi\over 2}{\left({\pi\xi\over 2}-1\right)}^{-1}
\right], &\hbox{if} \;\; {2\over\pi} < \xi \le {2\over\pi}{1\over 1-\epsilon}
\end{array} \right . \, .
\end{equation}

\n Note that $P(\xi,I)=0$ for $\xi > {2\over\pi}{1\over 1-\epsilon}$. The
above integral can be rewritten as
\begin{equation}
P(\xi,I) = {\pi E^0\over 2\alpha N_I}\int_1^l
{x(x-\epsilon)\over\sqrt{a+bx+cx^2}} \; dx 
\label{dbn1}
\end{equation}

\n where
\begin{equation}
a = - \epsilon^2 {\left({\pi\xi\over 2}\right)}^2,  \;\;
b = 2\epsilon {\left({\pi\xi\over 2}\right)}^2, \;\;
c = 1 - {\left({\pi\xi\over 2}\right)}^2 \, .
\end{equation}

\n This integral can be computed for any given value of the
parameters \cite{grad}.

\begin{figure}[h]
\centerline{\psfig{figure=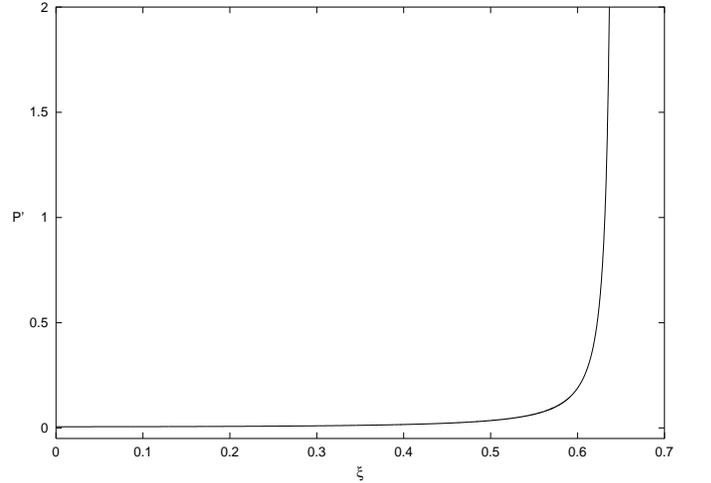,height=9cm,width=6.5cm,angle=-90}}
\caption{Typical behavior of the derivative $P'$ for $\xi<2/\pi$.
For $\xi>2/\pi$, $P'$ is not defined as the function $P$ is not smooth.}
\label{fig1}
\end{figure}

\section{results}

Using the above expression, it is possible to show that the derivative
\begin{equation}
P' = \left.{\partial
P\over\partial\alpha}\right|_{\alpha=\alpha_0}
\end{equation}

\n is positive for all values of  $\xi<2/\pi$. Moreover, $P'$, and hence the
sensitivity to $\alpha$, is maximal in the vicinity of the critical value
$\xi'=2/\pi$, as can be seen in Figure \ref{fig1}.

\begin{figure}[ht]
\centerline{\psfig{figure=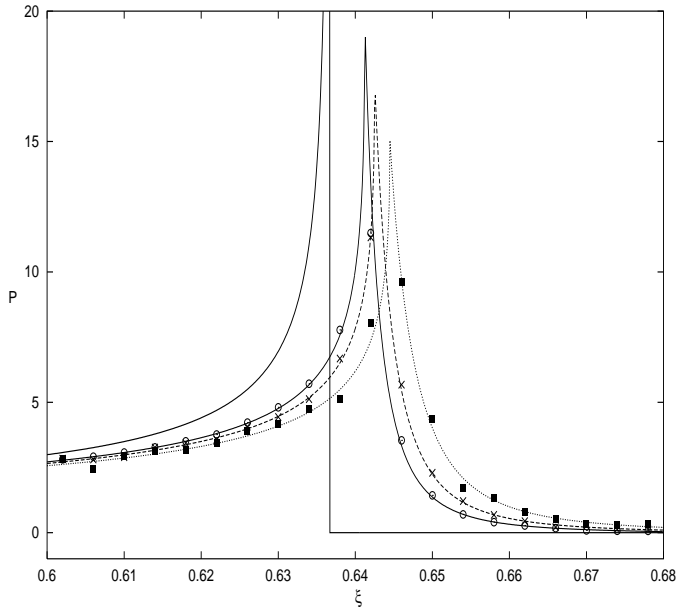,height=9cm,width=8cm,angle=-90}}
\caption{Nodal domain distribution for the rectangular boundary. Solid, dashed
and dotted curves are the approximate distribution (\ref{dbn1}) with $\alpha =
0.13,0.44,0.92$ respectively for the energy range $I=[10^2,10^4]$. This may be
compared with the limiting distribution (\ref{limit}) shown as a
thick curve.}
\label{fig2}
\end{figure}

In Figure \ref{fig2}, the nodal domain distribution given by the
eq. (\ref{dbn1}) is shown for different $\alpha$, along with the
corresponding numerical data. The limiting distribution is obtained by
taking the spectral interval to infinity. In this limit, $\epsilon=0$ and
\begin{equation}
P(\xi) = \left\{ \begin{array}{ll}
{\left[ 1 - {(\pi\xi/2)}^2  \right]}^{-1/2}, & \xi < 2/\pi \\[10pt]
0, & \xi > 2/\pi \end{array} \right .
\label{limit}
\end{equation}

\n which is independent of $\alpha$. Thus the parameter dependence is
arising from the leading finite energy correction to $P(\xi,I)$.

The problem studied above shows clearly that the nodal sequence stores
geometrical information, which, in the present case suffices to determine
unambiguously the rectangle for which the nodal sequence is given. Attempts
to generalize these ideas to other separable systems such as e.g., smooth
surfaces of revolutions or flat tori are under way. \\

\n {\bf Acknowledgments}

This research was supported in part  by the Minerva center for complex
systems and the Einstein (Minerva) center at the Weizmann Institute.
Grants from the German-Israeli Foundation and the Israel Science Foundation
are acknowledged.

\end{document}